\documentclass[%
reprint,
superscriptaddress,
nofootinbib,
amsmath,
amssymb,
aps,
floatfix,
showkeys,
]{revtex4-2}

\usepackage[normalem]{ulem}
\usepackage{fancyhdr}
\usepackage{graphicx,amsfonts,amssymb,amsbsy}
\usepackage{amsmath,amsthm,latexsym}
\usepackage[utf8]{inputenc}  
\usepackage{natbib}
\usepackage[colorlinks]{hyperref}
\usepackage{booktabs}
\usepackage{xcolor}

\hypersetup{linkcolor=magenta,citecolor=cyan}

\begin{document}

\title{Quark-hadron \textit{pasta} phase in neutron stars: \\ the role of medium-dependent surface and curvature tensions}

\author{Mauro Mariani}
\email{mmariani@fcaglp.unlp.edu.ar}

\affiliation{Facultad de Ciencias Astronómicas y Geofísicas, Universidad Nacional de La Plata, Paseo del Bosque S/N, 1900, Argentina}
\affiliation{CONICET, Godoy Cruz 2290, Buenos Aires, Argentina} 
\affiliation{Universidade Federal do ABC, Centro de Ciências Naturais e Humanas, Avenida dos Estados 5001- Bangú, CEP 09210-580, Santo André, SP, Brazil.}

\author{Germán Lugones}
\email{german.lugones@ufabc.edu.br}
\affiliation{Universidade Federal do ABC, Centro de Ciências Naturais e Humanas, Avenida dos Estados 5001- Bangú, CEP 09210-580, Santo André, SP, Brazil.}

\begin{abstract}
We investigate the properties of the hadron-quark mixed phase, often termed the  \textit{pasta} phase, expected to exist in the cores of massive neutron stars. To construct the equations of state (EoS), we combine an analytical representation based on the APR EoS for hadronic matter with the MIT bag model featuring vector interactions for quark matter. 
For modeling the mixed phase, we utilize the compressible liquid drop model that consistently accounts for finite-size and Coulomb effects.
Unlike most previous analyses that treated surface tension as a constant free parameter and neglected curvature tension, we employ microphysical calculations using the multiple reflection expansion formalism to determine these parameters, while also ensuring their self-consistency with the EoS.
We construct an extensive set of mixed hybrid EoSs by varying model parameters, solve the stellar structure equations to obtain neutron star mass-radius relationships, and select the models that satisfy current astrophysical constraints. Our findings closely align with calculations using a constant surface tension in terms of EoS stiffness and resulting stellar structure. However, they reveal significant differences in the types of geometric structures and their prevalence ranges within the mixed phase. Specifically, curvature effects enhance the emergence of tubes and bubbles at high densities despite the large value of surface tension, while suppressing the existence of drops and rods at low densities.
\end{abstract}

\maketitle

%-------------------------------------------------------------------
\section{Introduction}
\label{sec:intro}
%-------------------------------------------------------------------

Unveiling the intricate internal structure of neutron stars (NSs) has persistently remained an unresolved and challenging problem in astrophysics. While the comprehension of the equation of state (EoS) of matter within the crust and the outer core of NSs has, over time, reached an appreciable level of understanding, when it comes to investigating the behavior of matter in the inner core regions, where the density goes beyond twice the nuclear saturation density, we are faced with significant challenges. Guided by the current understanding of the Quantum Chromodynamics (QCD) phase diagram, we know that matter deconfines under extremely high-density conditions, suggesting that NS interiors could be described by the model of hybrid stars (HS). These astrophysical entities are envisaged as stellar bodies housing an inner core composed of quark matter surrounded by hadronic matter.  

The subject of HSs has been investigated for decades. The extensive study of these objects has indeed led to an advancement in our grasp of their general properties, but many issues are still difficult to pinpoint accurately. Among these challenges are understanding the precise structure of the quark-hadron interface, defining the exact state of matter under such extreme conditions, and determining the potential influences of these states on the macroscopic observable properties of NSs. 

In situations where the quark-hadron phase transition is assumed to be of the first order, the surface and curvature tensions, denoted by $\sigma$ and $\gamma$ respectively, stand out as crucial parameters. Numerous works (such as \cite{Heiselberg:1999mq, Voskresensky:2002csa, Maruyama:2007hqm,Xia:2019pnq, Yasutake:2019csp, Xia:2020brt}, among others) converge on the idea that depending on the value of the surface tension, the hadron-quark phase transition could be either a sharp discontinuity or a mixed phase where both phases coexist. It is generally accepted that if $\sigma$ falls below a critical value, $\sigma_\mathrm{crit}$, of the order of tens of MeV/fm$^2$, the energetically preferred state would be the mixed phase. Otherwise, a sharp boundary would exist between an inner core composed entirely of quark matter and the outer layers of pure hadronic matter. Many studies have assumed $\sigma$ to be either infinity or zero, leading to what are known as the Maxwell and Gibbs constructions, respectively.  In both approaches only bulk contributions to the system energy are considered. 
In scenarios where $\sigma$ is smaller than $\sigma_\mathrm{crit}$, the system's energy minimization favors global electrical neutrality over local neutrality. Consequently, a collection of electrically charged geometrical structures made of one phase arises within an electrically charged background of the other phase \cite{Ravenhall:1983uh, Glendenning:1992vb, Glendenning:2001pe, Endo:2006cse}. This outcome emerges from a finely balanced competition involving various finite-size factors, including surface, curvature, and Coulomb energies \cite{Haensel:2007nse}. The resulting mixed phase is often referred to as the quark-hadron \textit{pasta} phase, as the geometric structures (drops, rods, slabs, tubes, and bubbles), resemble Italian pasta varieties like gnocchi, spaghetti, and lasagna immersed in a uniform ``sauce''. 

Traditional analysis of the \textit{pasta} phase has primarily centered on understanding surface and Coulomb effects. However, several authors have underscored the potential significance of curvature energy \cite{Pethick:1983eft, Kolehmainen:1985tzw, Lorenz:1991dma, Thi:2021teo}. Guided by these insights, our current research will encompass not just the standard surface and Coulomb effects, but will also include contributions from curvature.

Despite the significance of surface and curvature tensions, their precise values in the dense environment of neutron stars are yet to reach a consensus. While we know that for hadronic matter $\sigma$ is relatively low (below $1 \mathrm{~MeV/fm}^2$) at the standard densities of a NS's crust, its determination at higher densities is reliant on phenomenological models, resulting in significant uncertainty. In this sense, it is generally posited that the surface tension of quark matter exceeds that of hadronic matter at such densities, being the decisive factor in the existence of the quark-hadron \textit{pasta} phase. Also, part of existing research has concentrated on the microphysical computation of $\sigma$, while the curvature tension has seen fewer investigations. All in all, the outcomes of these studies greatly depend on the specific EoS chosen and the method used to evaluate $\sigma$ and $\gamma$. For example, when using the thin-wall formalism, low surface tension values ($\sigma < 30~\mathrm{MeV}/\mathrm{fm}^{2}$) have been found across various EoSs, such as the NJL model \cite{Garcia:2013eaa, Ke:2013wga,Pinto:2012aq}, the linear sigma model \cite{Palhares:2010be, Pinto:2012aq, Kroff:2014qxa}, and the Polyakov-quark-meson model \cite{Mintz:2012mz}. Similarly, when the multiple reflection expansion (MRE) formalism is employed and the system is treated as a free particle gas, low values of $\sigma$ are also obtained \cite{Berger:1986ps,Lugones:2016ytl,Lugones:2018qgu,Lugones:2020qll}. However, when interactions are incorporated into the EoS, the MRE formalism tends to produce considerably higher $\sigma$ values. This pattern can be seen in studies using the NJL model \cite{Lugones:2013ema} and the vector MIT bag model \cite{Lugones:2021tee}. In contrast to surface tension, the study of curvature tension is less extensive. Nonetheless, in the context of the MRE formalism, light quarks have been observed to contribute minimally to $\sigma$ but significantly to $\gamma$. For more massive species, like the $s$ quark, both surface and curvature effects could be important \cite{Lugones:2021tee}. In an effort to sidestep the aforementioned uncertainties  and elaborate a clearer picture of potential mixed phases, numerous studies have opted to treat the surface tension of quark matter as a constant free parameter.

To provide a more comprehensive description of the \textit{pasta} phase, our approach will encompass the consideration of surface, Coulomb, and curvature effects adopting medium-dependent values for $\sigma$ and $\gamma$, and constructing the mixed phase within the framework of the compressible liquid-drop model (CLDM). The CLDM, which will be elaborated upon in subsequent sections, determines the equilibrium configuration of the mixed phase through the minimization of total energy density. Widely employed for investigating the nuclear liquid-gas phase transition at subnuclear densities \cite{Baym:1971nsm, Lattimer:1985ppo, Lorenz:1991dma, Bao:2014eot}, this method ensures thermodynamic consistency by incorporating finite-size energy contributions alongside bulk contributions within the minimization process. 
{This stands in contrast to the conventional coexisting phase approximation, which primarily solves for bulk equilibrium within the mixed phase (i.e., no finite size effects are incorporated in the mechanical and chemical equilibrium conditions) and incorporates finite-size effects in the energy density \cite{Maruyama:2005vb, Avancini:2008wac, Bao:2014iot}. Additionally, in Ref.~\cite{Avancini:2008wac}, the authors explore not only the coexisting phase approximation but also the self-consistent Thomas-Fermi approximation.}
While some recent studies in the literature have used the CLDM to model the hadron-quark mixed phase in neutron stars \cite{Wu:2017xaz, Wu:2019zoe, Ju:2021nev, Ju:2021hoy}, none of these have delved into the role of curvature energy or explored the incorporation of medium-dependent $\sigma$ and $\gamma$.

In recent years, gravitational wave detections and multi-messenger astronomy have provided crucial insights into neutron stars. Two important discoveries were PSR J1614-2230 \citep{Demorest:2010ats, Arzoumanian:2018tny} and PSR J0348+0432 \citep{Antoniadis:2013amp}, both observed to have around $2 M_\odot$. Another pulsar with a similar mass, PSR J0740+6620 \cite{Cromartie:2020rsd,Fonseca:2021rfa}, was later found. After these observations, any viable EoS must be capable of producing stellar configurations with a maximum mass above $2 M_\odot$. In addition, the NICER telescope, in collaboration with XMM-Newton, has provided valuable constraints on the mass and radius for PSR J0740+6620 \citep{Miller:2021tro,Riley:2021anv} and PSR J0030+0451 \citep{Miller:2019pjm,Riley2019anv}.  Finally, gravitational wave detectors have offered another avenue of crucial constraints via neutron star merger events.  To date, the LIGO-Virgo observatory has reported two significant events:  GW170817 \citep{Abbott:2017oog, Abbott:2018exr} and GW190425 \citep{Abbott:2020goo}. Notably, GW170817 was also detected in the electromagnetic spectrum, evidenced by events dubbed GRB170817A and AT2017gfo \citep{Abbott:2017gwa, Abbott:2017mmo}. These gravitational wave occurrences not only provide insights into the mass and radius but also shed light on the tidal deformability of these objects. We will use these findings to constrain the results of our study.

The work is structured as follows. In Section~\ref{sec:modelseos}, we present the  EoSs that we use to describe hadron and quark matter. In Section~\ref{sec:mixedphase}, we describe in detail the CLDM that we adopt to construct the mixed phase paying special attention to the incorporation of medium-dependent surface and curvature tensions. In Section~\ref{sec:results}, we present and analyze the results of our work, focusing on the microscopic geometric structures that compose the \textit{pasta} phase, as well as on the effects on the global properties of NSs. A summary of the work and a discussion of the implications of our results are presented in Section~\ref{sec:conclus}.

%-------------------------------------------------------------------
\section{Equation of state and finite size effects}
\label{sec:modelseos}
%-------------------------------------------------------------------

\subsection{Hadron matter EoS in bulk}
\label{sec:hadroneos}

For cold hadronic matter we use an analytic representation of the EoS of Akmal, Pandharipande and Ravenhall (APR) with boost corrections and three-body forces, $V_{18}+\delta v+$ UIX$^*$ \cite{Akmal:1998cf}. Hadronic matter is composed by protons, neutrons and leptons and the total energy per nucleon is written as \cite{Haensel:2007nse}:
\begin{align}
E = E_{\mathrm{N} 0} +   E_\mathrm{N} + E_l \, , 
\end{align}
where $E_{\mathrm{N} 0}$ is the nucleon rest-mass contribution, $E_\mathrm{N}$ is the energy per nucleon (excluding the rest-mass energy), and  $E_{l}$ is the lepton  contribution. The nucleon rest-mass contribution, is given by
\begin{equation}
    E_{\mathrm{N} 0}=\frac{\epsilon_{\mathrm{N} 0}}{n_B}=\frac{n_n m_n + n_p m_p}{n_B}=(1-x_p) m_n + x_p m_p \, ,
\end{equation}
where $\epsilon_{\mathrm{N} 0}$ is the rest-mass energy density, $n_B$ is the baryon number density, $n_n$ ($n_p$) is the neutron (proton) number density, $m_n$ ($m_p$) the neutron (proton) mass, and $x_p=n_p / n_B$ is the proton fraction. For $E_\mathrm{N}$ we will adopt a simple form consisting of a compressional term and a symmetry term:
\begin{align}
E_\mathrm{N}=W(u)+S(u)\left(1-2 x_p\right)^{2} \, ,
\end{align}
where $u \equiv n_B / n_{0}$ is the ratio of the baryon number density to the nuclear saturation density, $n_0=0.16$~fm$^{-3}$, $W(u)$ is the energy per nucleon in symmetric nuclear matter, and $S(u)$ is the symmetry energy. The relationship
\begin{equation}
\epsilon_\mathrm{N} = E_\mathrm{N} n_B,
\end{equation}
specifies the nucleon energy density used in the EoS.

A simple analytic representation of the compressional term $W(u)$ was constructed by Heiselberg and Hjorth-Jensen \cite{Heiselberg:1999mq}, which reproduces accurately the saturation density, the binding energy and the compressibility of the APR EoS:
\begin{equation}
W(u)=E_0 u(2+\delta-u) /(1+\delta u) \, . 
\end{equation}
By construction, $W(1)=E_{0}=$ $-15.8$~MeV, in accordance with the experimental value. The free parameter $\delta$ is related to the incompressibility of symmetric nuclear matter at the saturation point \cite{Haensel:2007nse},
\begin{align}
K_0=9\left(\frac{\mathrm{d}^2 W}{\mathrm{~d} u^2}\right)_{u=1}=\frac{18\left|E_0\right|}{1+\delta} \, .
\end{align}
In the following, we will consider three values for $\delta$, namely $\delta=0.1, 0.15,0.2$. By adopting the value $\delta = 0.1$,  the corresponding nuclear incompressibility constant is $K_0=258$~MeV; when we adopt  $\delta = 0.15$, we obtain $K_0=247$~MeV; and for $\delta = 0.2$, we obtain $K_0=237$~MeV. All of these values are in agreement with the experimental results derived from the analysis of isoscalar giant monopole resonances in heavy nuclei, $K_0=240 \pm 20 \mathrm{MeV}$ \cite{Colo:2004mj,Garg:2018uam,Kumar:2021vdr,Perego:2021mkd}.

For the symmetry energy $S(u)$, the APR results can be fitted with a simple formula \cite{Heiselberg:1999mq}:
\begin{align}
S(u)=32 u^{\zeta} \mathrm{MeV} \, ,
\end{align}
which results in the following expression for the derivative $L$ of the symmetry energy:
\begin{align}
L \equiv 3 \frac{d S(u)}{d u}  = 96 \zeta u^{\zeta-1} \mathrm{MeV} \, . 
\end{align}
Using $\zeta=0.6$ \cite{Heiselberg:1999mq} one obtains  $S_0 \equiv S(1)  = 32$~MeV and $L_0 \equiv L(1) = 58$~MeV, which are in agreement with experimental values \cite{Burgio:2020fom}. 

Regarding the lepton contribution, we take into account both electrons and muons, i.e. $E_l = E_e + E_\mu$. Each lepton species is incorporated as a free Fermi gas, with the pressure given by
\begin{eqnarray}
    P_e & =& \frac{\mu^4_e}{12\pi^2} \, , \label{eq:leppressure} \\
    P_\mu &= & \frac{1}{12\pi^2} \bigg[\mu_\mu \nu_\mu (\mu^2_\mu-\tfrac{5}{2} m^2_\mu) \nonumber \\
          & & \qquad + \tfrac{3}{2} m^4_\mu \ln \left( \frac{\mu_\mu+\nu_\mu}{m_\mu} \right) \bigg] \, ,
\end{eqnarray} 
and the particle number density by,
\begin{gather}
    n_e = \frac{\mu_e^3}{3 \pi^2} \, , \\
    n_\mu = \frac{\nu_\mu^3}{3 \pi^2} \, , \label{eq:lepnumber}
\end{gather}
where $\nu_\mu=(\mu^2_\mu-m^2_\mu)^{1/2}$, $\mu_e$ and $\mu_\mu$ are the electron and muon chemical potential,  and $m_\mu=105.7$~MeV is the muon mass. Due to chemical equilibrium and the free escape of neutrinos from the system, we have $\mu_e = \mu_\mu$. The energy density for each lepton $l$ can be obtained through the Euler relationship,
\begin{equation}
    \epsilon_l = - P_l + \mu_l n_l \, . \label{eq:eulerlepton}
\end{equation}

Finally, we find that when $\delta = 0.10$ and $\zeta = 0.6$, the EoS becomes superluminal at densities of the order of $11 n_0$. For the cases of $\delta = 0.15, 0.2$ and $\zeta = 0.6$, the EoSs remain subluminal even at significantly higher densities. 

In summary, our analytic model approximates the APR EoS \cite{Akmal:1998cf} considering neutron, protons, electrons and muons, it aligns with experimental data, and displays the appropriate causal behavior across the entire range of densities relevant to NSs. For pure hadronic matter, the model is supplemented with the conditions of electric charge neutrality and $\beta$-equilibrium.

{To conclude, it is worthwhile to add some remarks on the reliability of the results that will be obtained using the previously presented EoS. Despite its schematic nature, the EoS aligns with all relevant empirical data at nuclear saturation density. It is crucial to acknowledge, however, that any description of hadronic matter at densities exceeding nuclear saturation essentially constitutes an extrapolation.  While various models in the literature provide more sophisticated approaches to nuclear interactions,  they too face substantial uncertainties, arising from factors such as the selection of parameters and/or the specific terms included in the Lagrangian for modeling nuclear interactions. In this context, the relative simplicity of our chosen model does not necessarily detract from the accuracy of the resultant EoS in mirroring the true EoS of actual NSs, compared to other phenomenological models. This is primarily because robust constraints on the extrapolations made by these models  are currently  quite broad and still not very restrictive. Such criteria include thermodynamic consistency, causality,  alignment with results from perturbative QCD at ultra-high densities, and astrophysical constraints. Of course, it is extremely important that future works extend the current calculations by incorporating other EoS to gain a broader understanding of the range of possible outcomes. A more detailed description of the effects of the EoS choice on the main results of this work is provided in the conclusions section.}

%-------------------------------------------------------------------
\subsection{Quark matter EoS in bulk}
\label{sec:quarkeos}
%-------------------------------------------------------------------

Quark matter is described by the MIT bag model with vector interactions, which are introduced by a vector-isoscalar meson $V^{\mu}$ with a universal coupling constant $g$, coupled to all three quarks, $u$, $d$, and $s$. The Lagrangian density of the model reads \cite{Franzon:2016urz,Lopes2021_I,Lopes2021_II}:
\begin{equation}
\begin{aligned}
\mathcal{L} & = \sum_{q= u, d, s}\left\{\bar{\psi}_q\left[i \gamma^{\mu} \partial_{\mu}-m_q\right] \psi_q-B\right\} \Theta\left(\bar{\psi}_q \psi_q\right)  \\
& + \sum_{q=u, d, s} g \left\{\bar{\psi}_q \gamma^{\mu} V_{\mu} \psi_q\right\} \Theta\left(\bar{\psi}_q \psi_q\right) +   \tfrac{1}{2} m_V^{2} V_{\mu} V^{\mu} \\
& + \sum_{l = e, \mu}  \bar{\psi}_{l} \left(i \gamma_{\mu} \partial^{\mu}-m_{l}\right) \psi_{l}  \, ,
\end{aligned}
\end{equation}
where $q$ runs over quarks and $l$ over leptons,  the bag constant $B$ represents the extra energy per unit  volume required to create a region of perturbative vacuum \cite{Farhi:1984qu}, and  $\Theta$ is the Heaviside step function  ($\Theta=1$ inside the bag, $\Theta=0$ outside).  
The mass of the vector field is taken to be $m_V = 780 \; \mathrm{MeV}$.

Working in the mean field approximation and defining $G_{V} \equiv \left({g}/{m_V}\right)^{2}$, the eigen-energy of the quarks reads
\begin{equation}
E_q   = \sqrt{m_q^2+k^2}+ G_V^{1/2} m_V V_0\, ,  
\end{equation}
being $k$ the particle's momentum and $m_q$ the quark mass ($m_u = 2.16~\mathrm{MeV}$, $m_d = 4.67~\mathrm{MeV}$, $m_s = 93.4~\mathrm{MeV}$). The equation for the mean vector field $V_0$ is given by: 
\begin{equation}
m_V V_0  = G_V^{1/2} (n_u + n_d + n_s) \, ,  
\end{equation}
where $n_q=\left\langle\bar{\psi}_q \gamma^{0} \psi_q\right\rangle$ is the  quark number density. 
The grand thermodynamic potential per unit volume at zero temperature is \cite{Lopes2021_II}:
\begin{eqnarray}
\Omega & =&\sum_{i = q, l} \Omega^*_{i} + B-\tfrac{1}{2} m_V^{2} V_0^{2} \, ,
\label{eq:Omega_bulk}
\end{eqnarray}
where for quarks,
\begin{equation}
\Omega^*_q= - \frac{g_q}{6 \pi^{2}} \int_{0}^{k_{Fq}} \frac{k^4 d k}{\sqrt{k^2 + m_q ^2}}  \, , 
\end{equation}
being  $g_q=6$ the degeneracy factor. The Fermi momentum, $k_{Fq}$, is given by
\begin{equation}
k_{Fq}  =(\mu_{i}^{*2} - m_i ^2)^{1/2},
\end{equation}
where the effective chemical potential for quarks reads
\begin{eqnarray}
\mu_q^{*} &=&  \mu_q -  G_V^{1/2} m_V V_0 \, ,
\label{eq:effective_mu}
\end{eqnarray}
being $\mu_q$ the chemical potential of quarks of flavor $q$.
The particle number density of each quark species is 
\begin{eqnarray}
n_q =  \frac{g_q}{2 \pi^2} \int_{0}^{k_{Fq}} k^2 dk \, ,
\end{eqnarray}
and the energy density,
\begin{eqnarray}
\epsilon_q =  \frac{g_q}{2 \pi^2} \int_{0}^{k_{Fq}} E_q k^2 dk \, .
\end{eqnarray}

{Within this quark EoS model,  the speed of sound ($c_s$) depends on $G_V$ and is not affected by the bag constant. For $G_V = 0$, the speed of sound remains constant at $(c_s/c)^2 = 1/3$. However, when $G_V>0$, $c_s$ increases with density, asymptotically approaching the causal limit where $c_s = c$. Furthermore, at a fixed $n_B$, a larger $G_V$ results in a higher $c_s$. The  values of $c_s$ arising in our calculations always fall within the range of $1/3 < (c_s/c)^2 < 0.6$. This upper limit of $(c_s/c)^2 \approx 0.6$ occurs when using the maximum $G_V$ value of $0.3\, \mathrm{fm}^2$ and the highest density ($n_B/n_0 \sim 11$).}

Leptons -electrons and muons- are incorporated as free Fermi gases, following the formulas specified in Eqs.~\eqref{eq:leppressure}-\eqref{eq:lepnumber}.  The construction of the pure quark EoS is carried out under the conditions of electric charge neutrality and chemical equilibrium under weak interactions.

{Finally, it is worth briefly commenting on the impact of choosing a specific model among all the available phenomenological models to describe quark matter. Clearly, the use of alternative quark models, such as the Nambu-Jona-Lasinio model or the linear sigma model, would lead to different results. These variations would be seen in the transition pressure, the stiffness of the EoS, and the quark concentrations at each density. In this respect, our results are quite dependent on the chosen model, as are many other outcomes related to NS interiors. However, as we will discuss more thoroughly in Sec. \ref{sec:conclus}, the fundamental conclusion of our paper hinges on the qualitative behavior of the density-dependent surface and curvature tensions. In this context, we will conclude that, despite the current uncertainties in EoS, some robust conclusions can be drawn about the behavior of the mixed phase in NSs.}

%-------------------------------------------------------------------
\section{Mixed phase within the compressible liquid drop model}
\label{sec:mixedphase}
%-------------------------------------------------------------------

To investigate the mixed phase of hadrons and quarks, we employ a technique similar to the compressible liquid-drop model (CLDM) used for examining nuclear liquid-gas phase transitions at subnuclear densities \cite{Baym:1971nsm, Lorenz:1991dma}. This approach enables a thermodynamically consistent and systematic analysis of both bulk and finite-size effects, making it particularly suitable for studying phase transitions in dense matter. We apply the model within the framework of the Wigner-Seitz (WS) approximation, which divides the entire space into independent equivalent cells characterized by their geometric symmetry. 

{Each WS cell is assumed to be electric charge-neutral. This is a reasonable assumption since, as with any plasma, there is no significant charge separation on scales larger than the Debye screening length $\lambda_D$. For dense quark matter, $\lambda_D \sim$ few fermis \cite{Heiselberg:1993qmd}, in agreement with the size of the WS cells resulting from the CLDM calculations.} 

{As we work under the hypothesis of a first order hadron-quark phase transition,
the coexisting hadronic and quark phases are assumed to be separated by a sharp boundary, with uniform particle densities in each phase for the sake of simplicity. A priori, the assumption that charged baryons and leptons are uniformly distributed is a simplification.  A more realistic approach would involve solving the Poisson equation for the electrostatic potential self-consistently, as in Refs. \cite{Yasutake:2014oxa, Schmitt:2020tac}. This method would more accurately reflect the local variations in charge distribution and screening effects. Despite this, our  approach remains quite realistic,  as can be checked in Fig. 4 of \cite{Yasutake:2014oxa}. First, the distribution of charged baryons is found to be almost uniform within each phase. On the other hand, leptons are not as uniformly distributed. This aspect warrants further refinement in future analyses. However, the approximation used here is still reasonably accurate since electrons hold a very small fraction of the total charge and are smoothly distributed \cite{Yasutake:2014oxa}.}

%-----------------------------------------------
\subsection{Energy density of the mixed phase}
%-----------------------------------------------

The energy density of the hybrid mixed phase is expressed as \cite{Haensel:2007nse}:
\begin{equation}
    \epsilon_\mathrm{MP} = \epsilon_{H, \mathrm{bulk}} + \epsilon_{Q, \mathrm{bulk}} + \epsilon_l + \epsilon_\mathrm{surf} + \epsilon_\mathrm{curv} + \epsilon_\mathrm{Coul} \, ,
    \label{eq:eps}
\end{equation}
where $\epsilon_{H, \mathrm{bulk}}$ represents the bulk contribution of hadrons to the energy density, $\epsilon_{Q, \mathrm{bulk}}$ denotes the quark bulk contribution, and $\epsilon_l$ refers to the common lepton background (described as a free Fermi gas of electrons and muons, as shown in Eqs.~\eqref{eq:leppressure}-\eqref{eq:lepnumber}). Additionally, $\epsilon_\mathrm{surf}$, $\epsilon_\mathrm{curv}$, and $\epsilon_\mathrm{Coul}$ correspond to the surface, curvature, and Coulomb contributions, respectively.

The bulk energy density of hadrons and quarks are, respectively:
\begin{eqnarray}
   \epsilon_{H, \mathrm{bulk}} &=& (1 - \chi)  \epsilon_H  , \\
   \epsilon_{Q, \mathrm{bulk}} &=& \chi  \epsilon_Q ,
\end{eqnarray}
where $\epsilon_H$ and $\epsilon_Q$ are the uniform energy densities within each phase inside the WS cell. The volume fraction is defined as: 
\begin{equation}
\chi = \frac{V_d}{V_{\mathrm{WS}} } = \left(\frac{r_d}{r_{\mathrm{WS}}} \right)^d  \\
\label{eq:chi}
\end{equation}
where $V_d$ and $r_d$ represent the volume and size of each geometric structure. Similarly, $V_{\mathrm{WS}}$ and $r_{\mathrm{WS}}$ correspond to the volume and size of the WS cell, and $d$ denotes the dimensionality of the geometric structure (with $d=1$ for slabs, $d=2$ for rods and tubes, and $d=3$ for drops and bubbles).

The finite size effects contributions are given by \cite{Ravenhall:1983uh, Lorenz:1991dma}:
\begin{eqnarray}
    \epsilon_\mathrm{surf} &=& \frac{d \sigma \chi_\mathrm{in}}{r_d} \,, \label{eq:eps_surf}\\
    \epsilon_\mathrm{curv} &=& \pm \frac{d (d-1) \gamma \chi_\mathrm{in}}{r_d^2} \,, \label{eq:eps_curv}\\
    \epsilon_\mathrm{Coul} &=& \frac{1}{2} \delta q^2 r_d^2 \chi_\mathrm{in} f_d(\chi_\mathrm{in}) \, ,   \label{eq:eps_coul}
\end{eqnarray}
where $\sigma$ and $\gamma$ represent the surface and curvature tensions, respectively. In the curvature energy term, $\epsilon_\mathrm{curv}$, the $+$ sign is assigned to droplets and rods, and the $-$ sign pertains to tubes and bubbles \cite{Pethick:1983eft, Lorenz:1991dma}. Meanwhile, the slab configuration lacks a curvature contribution ($d=1$).
The quantity $\chi_\mathrm{in}$ denotes the volume fraction of the inner portion,
\begin{equation}
    \chi_\mathrm{in}=
        \begin{cases}
            \chi \,, &\textrm{for } 0 < \chi < 0.5 \, , \\
            1-\chi \,, &\textrm{for } 0.5 < \chi < 1 \, . \\        
        \end{cases}
\end{equation}
The electric charge-density difference, $\delta q$, between the two phases inside the cell is calculated as $\delta q = e \delta n_c$, where $\delta n_c=n_{c,H}-n_{c,Q}$ represents the difference in the number of charged particles between the hadron and quark phases, being
\begin{eqnarray}
    n_{c,H} &=& n_p \, , \\
    n_{c,Q} &=& \tfrac{2}{3}n_u - \tfrac{1}{3}n_d - \tfrac{1}{3}n_s \, .
\end{eqnarray}
The elementary charge is denoted by $e=\sqrt{4 \pi/137}$, since we are operating within the Natural SI unit system.
The function $f_d$ in the Coulomb energy depends on the value of $d$ as follows:
\begin{equation}
    f_d(x)=
        \begin{cases}
            \dfrac{1}{d+2} \left( \dfrac{2-d x^{1-2/d}}{d-2} + x \right), &\textrm{if } d=1,3 \, , \\
            \dfrac{x-1- \ln(x)}{d+2}  \, , &\textrm{if } d=2 \, .
        \end{cases}
\end{equation}

%-----------------------------------------------
\subsection{Surface and curvature tensions}
\label{sec:surf_and_curv}
%-----------------------------------------------

As mentioned in the Introduction, no widely accepted consensus exists on the values of $\sigma$ and $\gamma$ needed to calculate Eqs.~\eqref{eq:eps_surf} and \eqref{eq:eps_curv}. To address these uncertainties, the majority of studies take certain simplifications into account. Firstly, they choose to neglect the role of $\gamma$. Secondly, the surface tension of hadrons is often disregarded, as it is assumed to be smaller than that of quark matter. Lastly, quark matter's surface tension is typically treated as a constant free parameter, providing flexibility in exploring various scenarios (see e.g. \cite{Yasutake:2014oxa, Maslov:2018ghi, Wu:2019zoe, Ju:2021nev} and references therein).

In this work, we adopt a different approach. For simplicity, we continue neglecting the contributions of hadrons to $\sigma$ and $\gamma$. However, for quark matter, instead of assuming constant values or neglecting these factors, we rely on microscopic calculations based on the MRE formalism \cite{Lugones:2013ema,Lugones:2021tee,Lugones:2020qll}.  Specifically, we use the results of Ref.~\cite{Lugones:2021tee}, which provide $\sigma$ and $\gamma$ for the same MIT bag model with vector interactions used in this work (see Sec. \ref{sec:quarkeos}). This approach allows us to determine $\sigma$ and $\gamma$ self-consistently with the EoS.  

The results of Ref.~\cite{Lugones:2021tee} reveal that $\sigma$ and $\gamma$ at zero temperature show a slight dependence on the electric charge and the geometric structure's size, but exhibit strong dependence on the density. Additionally, the presence of repulsive vector interactions significantly enhances $\sigma$ and $\gamma$. To simplify the description,  we assume that, for a given value of $G_V$, $\sigma$ and $\gamma$ solely depend on the baryon number density $n_B$. Based on data extracted from Ref.~\cite{Lugones:2021tee}, we construct functions of the form  $\sigma=\sigma(n_B, G_V)$ and $\gamma=\gamma(n_B, G_V)$. These functions will be presented subsequently along with our computational outcomes.

%-----------------------------------------------
\subsection{Minimization of the energy}
%-----------------------------------------------

The equilibrium of the system can be determined by minimizing the total energy density, {Eq.~\eqref{eq:eps}, with respect to the independent variables of the system, while considering baryon number conservation and global charge neutrality. Initially, for a given value of $n_B$, the variables are $\chi$, $r_d$, $n_p$, $n_n$, $n_u$, $n_d$, $n_s$, and $n_e$ but with the inclusion of these two constraints the number of independent variables can be reduced. Baryon number conservation and global charge neutrality can be, respectively,} expressed as:
\begin{gather}
\frac{\chi}{3}\left(n_{u}+n_{d}+n_{s}\right)+(1-\chi)\left(n_p+n_n\right)=n_{B} \, , \label{eq:nb} \\
\chi n_{c,Q}+(1-\chi) n_{c,H} - n_e - n_\mu = 0 \, . \label{eq:charge}
\end{gather}
The latter two equations can be employed to eliminate the variables $n_e$ and $n_n$. Consequently, for a given baryon number density $n_{B}$, the energy density $\epsilon_\mathrm{MP}$ still depends on six variables: $\chi$, $r_d$, $n_p$, $n_{u}$, $n_{d}$, and $n_{s}$.

Minimization with respect to changes in the size of the inner phase $r_d$, $\partial \epsilon_\mathrm{MP}/\partial r_d = 0$, while maintaining constant values for $\chi$ and all densities, determines the equilibrium size of the cell. The result is the standard ``virial theorem'':
\begin{align}
2 \epsilon_\mathrm{Coul} = \epsilon_\mathrm{surf} + 2 \epsilon_\mathrm{curv}    \, .
\end{align}
{From the latter expression we derive a quartic equation by employing Eqs. \eqref{eq:eps_surf}-\eqref{eq:eps_coul}. This equation is then solved to ascertain the value of $r_d$. Subsequently, utilizing Eq. \eqref{eq:chi}, we determine the value of $r_{\mathrm{WS}}$.}

%-------------------
\begin{figure*}[tbh]
\centering
\includegraphics[width=.99\linewidth]{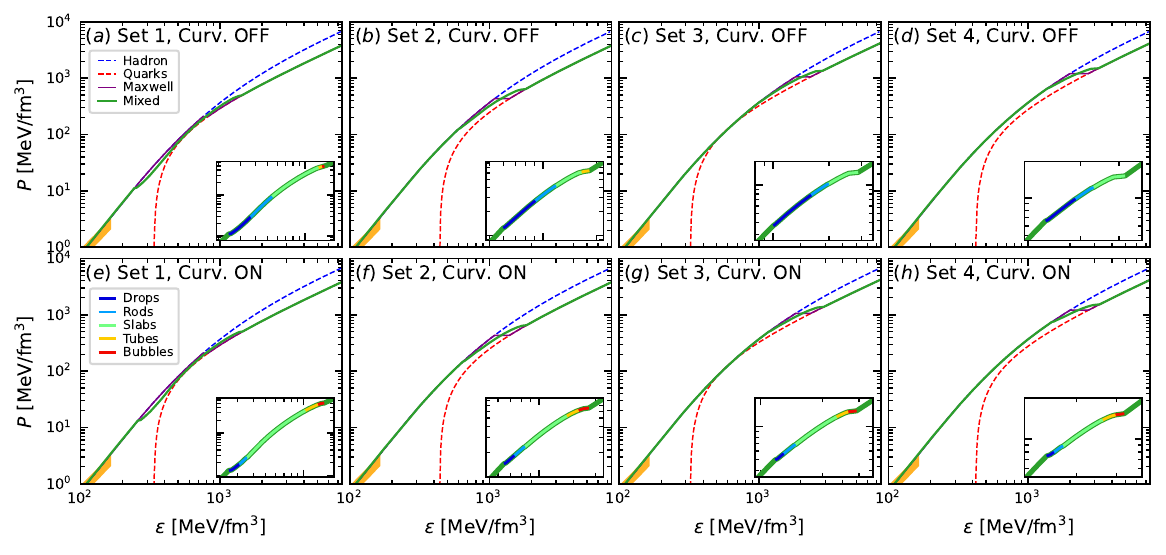}
\caption{Pressure-energy density relationship for the selected EoS sets in the absence (upper row) and presence (lower row) of curvature effects. Each panel shows a different EoS set, with the mixed EoS displayed in green, the Maxwell's construction EoS shown in purple and the pure hadron and quark phases represented by the dashed blue and red curves, respectively. Zoom boxes provide a closer look at the mixed region, highlighting the distinct geometrical structures with different colors. The orange region in the bottom-left corner represents the constraint given by chiral effective field theory up to $1.1\,n_0$ \citep{Hebeler:2013nza, Annala:2020efq}, which is satisfied by our hadronic EoS. In the cases without curvature effects, tubes and bubbles either do not form or are only marginally present. Conversely, when curvature effects are taken into account, the formation of tubes and bubbles is enhanced. At the same time, there is a reduction of the density extension of drops and rods, while the prevalence of slabs increases.}
\label{fig:eos}
\end{figure*}
%-------------------

Minimizing with respect to the volume fraction $\chi$, $\partial \epsilon_\mathrm{MP}/\partial \chi = 0$, leads to the mechanical equilibrium condition at the interface:
\begin{equation}
\begin{aligned} 
P_{H}= & P_{Q}-\frac{2 \epsilon_{\mathrm{Coul}}}{\delta q}\left[\frac{ \tfrac{2}{3} n_u - \tfrac{1}{3} n_d - \tfrac{1}{3} n_s}{ \chi} + \frac{n_p}{1-\chi} \right] \\ & \mp \frac{\epsilon_{\mathrm{surf}}+\epsilon_{\mathrm{curv}}}{\chi_\mathrm{in}} \mp \frac{\epsilon_{\mathrm{Coul}}}{\chi_\mathrm{in}}\left(1+\chi_\mathrm{in} \frac{f_d^{\prime}}{f_d}\right)     \, ,
\end{aligned}    
\end{equation}
where, in the last two terms, the $-$ sign applies  for droplets, rods, and slabs configurations, and the $+$ sign is used for tubes and bubbles. The total pressure of the mixed phase, $P_\mathrm{MP}$, is obtained by the thermodynamic relation,
\begin{equation}
P_\mathrm{MP} = n_B^2 \frac{ \partial \left( \epsilon_\mathrm{MP} / n_B \right) }{\partial n_B } \, .
\end{equation}

Finally, minimization with respect to the particle number densities $n_p$, $n_{u}$, $n_{d}$, and $n_{s}$, $\partial \epsilon_\mathrm{MP}/\partial n_i = 0$, results in the following chemical equilibrium conditions: 
\begin{eqnarray}
\mu_{p}+\frac{2 \epsilon_{\mathrm{Coul}}}{(1 - \chi) \delta q} & = &  \mu_{n}-\mu_{e} \, , \\ 
\mu_{u}-\frac{4 \epsilon_{\mathrm{Coul}}}{3 \chi  \delta q} & = & \tfrac{1}{3} \mu_{n}-\tfrac{2}{3} \mu_{e} \,, \\ 
\mu_{d}+\frac{2 \epsilon_{\mathrm{Coul}}}{3 \chi \delta q} & = & \tfrac{1}{3} \mu_{n}+\tfrac{1}{3} \mu_{e} \,, \\ 
\mu_{s}+\frac{2 \epsilon_{\mathrm{Coul}}}{3 \chi \delta q} & = &  \tfrac{1}{3} \mu_{n}+\tfrac{1}{3} \mu_{e} \,   , \label{eq:mue}
\end{eqnarray}
{which are the traditional $\beta$-equilibrium equations modified by an extra term, due to the Coulomb energy density contribution.}

{The mixed phase, with bulk and finite-size contributions simultaneously and consistently treated, is determined by solving the set of equations consisting of Eqs.~\eqref{eq:nb}-\eqref{eq:mue}. These equations are resolved for various fixed values of $n_B$ and for each dimensionality $d=1,2,3$. For each value of $n_B$, this procedure results in three distinct solutions, each corresponding to one of the $d$ values. Additionally, the system is resolved independently for each of the two pure phases -- hadrons and quarks -- resulting in a total of five different EoSs. Constructing the final mixed EoS involves comparing these five EoSs. For every $n_B$ value, the EoS that exhibits the lowest energy density is considered the prevailing one. An alternative methodology, to be discussed in Sec. \ref{sec:sigma_variable}, entails computing all the aforementioned phases at a given pressure $P$. At each pressure point $P$, the phase that minimizes the Gibbs free energy per baryon, $G$, is identified as the prevailing one.}

%-------------------------------------------------------------------
\section{Results}
\label{sec:results}
%-------------------------------------------------------------------

In this section, we provide a thorough analysis of the quark-hadron mixed phase. This analysis is built upon the EoSs detailed in Sec. \ref{sec:modelseos}, and it employs the theoretical framework outlined in Sec. \ref{sec:mixedphase}. Subsection~\ref{sec:sigma_variable} is dedicated to exploring the mixed-phase while taking into account medium-dependent surface and curvature tensions. Our focus extends beyond merely assessing the stiffness of the EoS; we also delve into the intricate properties of the microscopic geometric structures intrinsic to the \textit{pasta} phase. Additionally, we undertake an examination of the mass-radius relationship for the resulting NSs.  Subsection \ref{sec:sigmacte} is devoted to the constant $\sigma$ scenario. This subsection aims to draw comparisons between results obtained from this widely used approximation and our comprehensive study, which incorporates  density-dependent $\sigma$ and $\gamma$.

\begin{table}[tb]
\begin{tabular}{cccc}
\toprule
Set & $\delta$ & $G_V$ [fm$^{2}$] & $B$ [MeV/fm$^{3}$] \\
\midrule
$1$ & $0.10$ & $0.2$ & $90$ \\
$2$ & $0.10$ & $0.2$ & $120$ \\
$3$ & $0.10$ & $0.3$ & $90$ \\
$4$ & $0.10$ & $0.3$ & $120$ \\
$5$ & $0.15$ & $0.3$ & $90$ \\
$6$ & $0.15$ & $0.3$ & $120$ \\
\bottomrule
\end{tabular}
\caption{Parameter values of the filtered EoS sets; for all the sets we take $\zeta=0.6$. These sets are selected in order to satisfy the current astrophysical constraints, see text for details.}
\label{table:sets}
\end{table}

\begin{figure}[tb]
\centering
\includegraphics[width=0.75\linewidth]{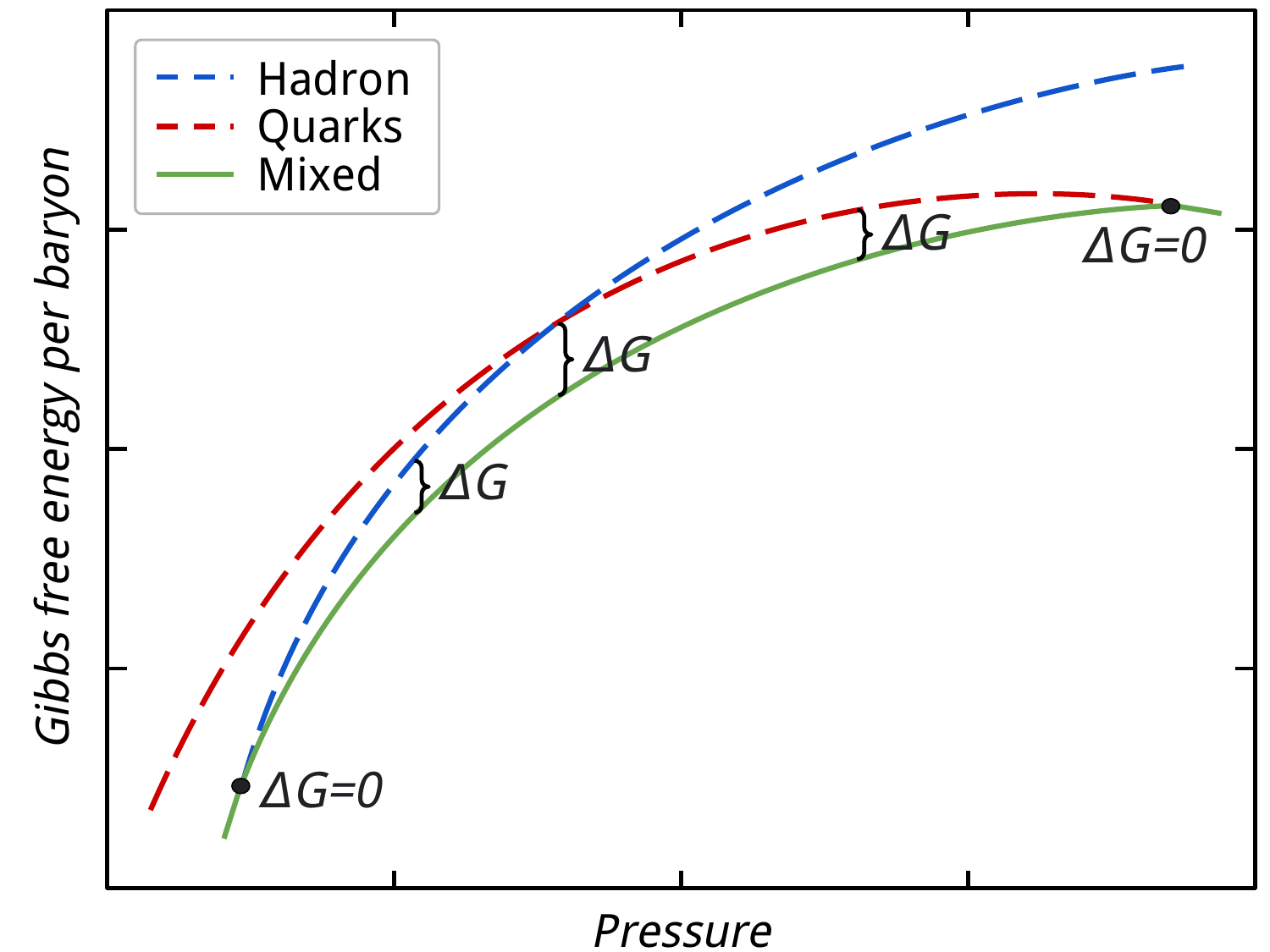}
\caption{Schematic representation of the mixed phase construction process. The green curve represents the EoS of a nonspecific geometric structure of the pasta phase, while the dashed blue and dashed red curves correspond to the pure hadron and pure quark EoSs, respectively. The $\Delta G$ magnitude is defined in Eq.~\eqref{eq:delta_G_definition}. For clarity, the plot presents the curve for only a single geometric structure. In practice, the value of $\Delta G$ is calculated for all geometric structures, and the one with the largest modulus, $|\Delta G|$, is energetically preferred. The start and end of the mixed phase are indicated by $\Delta G = 0$. To illustrate the determination of this magnitude, different values of $\Delta G$ are displayed at various pressures.}
\label{fig:delta_gibbs_construction}
\end{figure}

%---------------------------------------------------
\subsection{Analysis of the mixed phase incorporating medium-dependent surface and curvature tensions}
\label{sec:sigma_variable}
%---------------------------------------------------

The EoS for the mixed phase contains four free parameters, two from the hadron sector, $\delta$ and $\zeta$, and two from the quark sector, $G_V$ and $B$. For our calculations, we adopt the following parameter values:
 \begin{eqnarray}
    \delta &=& 0.1,\, 0.15,\, 0.2\, , \\
    \zeta &=& 0.6 \, , \\
    G_V &=& 0.1,\, 0.2,\, 0.3 \ \mathrm{fm}^{2} \, , \\
    B &=& 90,\, 120 \ \mathrm{MeV/fm}^{3} \, .  
 \end{eqnarray}
These values are aligned with previous works that develop and/or adopt the EoSs we use, and are chosen in order to cover a wide and representative range in the parameter space \cite{Lopes2021_I, Lopes2021_II,Lugones:2021tee}. Using the $18$ possible combinations of these values, we calculate the mixed phase EoSs, with an option to include or exclude the curvature effect. Additionally,  for comparison purposes, we  compute a hybrid EoS with a sharp density discontinuity using the Maxwell's construction. We integrate the Tolman-Oppenheimer-Volkoff (TOV) equations for all these $18$ EoSs and filter them in order to retain only those that meet the existing astrophysical constraints. Based on the outcomes from these models, we find that the most stringent constraint is the one requiring a maximum mass of $2 M_\odot$ \cite{Fonseca:2021rfa}. Therefore, any EoS failing to achieve this value is excluded. Details regarding these filtered EoSs are displayed in Table~\ref{table:sets}. Out of the filtered EoSs detailed in Table~\ref{table:sets}, we only present figures for the first four sets.  These sets are qualitatively representative of our results, aiding in the elucidation of the key aspects of our research. Also, for sets~$1$ and $2$, we present in Subsection~\ref{sec:sigmacte}, a detailed study of the widely used constant $\sigma$ scenario, in order to compare with the results of our study featuring a medium-dependent surface tension.

In Fig.~\ref{fig:eos}, we depict the pressure-energy density relationship for the chosen EoS sets in two scenarios: one without curvature effects (upper row) and the other with curvature effects (lower row). In addition to the mixed phase curve featuring several geometric structures, we show, for comparison, the pure hadronic and quark EoSs, as well as the EoS obtained through the Maxwell's construction method. The zoom boxes provide a detailed view of the presence and extension of each \textit{pasta} structure. 
Sets 1 and 2, both with $G_V=0.2$~fm$^2$, exhibit broad mixed phases. Set~$1$, in particular, has a notably wide mixed phase, beginning at a low pressure around $P \sim 10$~MeV/fm$^3$ and extending up to approximately $P \sim 500$~MeV/fm$^3$. This characteristic implies a significant softening of the mixed EoS when compared to the scenarios involving the Maxwell's construction or the pure hadronic EoS. For sets~$3$ and $4$, the higher $G_V$ values shift the transition to higher densities and results in a less extended mixed phase, approaching the behavior of the Maxwell's construction EoS.

Upon comparing both rows of Fig.~\ref{fig:eos}, it becomes evident that the inclusion of curvature effects leads to a higher prevalence of slabs over drops and rods, while also favoring the appearance of tubes and bubbles. This is due to the curvature effects, which reduce the energy associated with these specific structures (cf. the sign switching in Eq.~\eqref{eq:eps_curv}). While without curvature effects only set~$1$ manages to form bubbles, when including the curvature energy all sets exhibit bubble appearance. The impact of curvature on the range of the whole mixed phase is relatively minor: it increases the energy for drops and reduces it for bubbles, producing a subtle effect on the start and end of the mixed phase, shifting it towards slightly higher values of pressure and energy density. These changes, however, are not readily discernible in the figures.

%-------------------
\begin{figure*}[tbh]
\centering
\includegraphics[width=.99\linewidth]{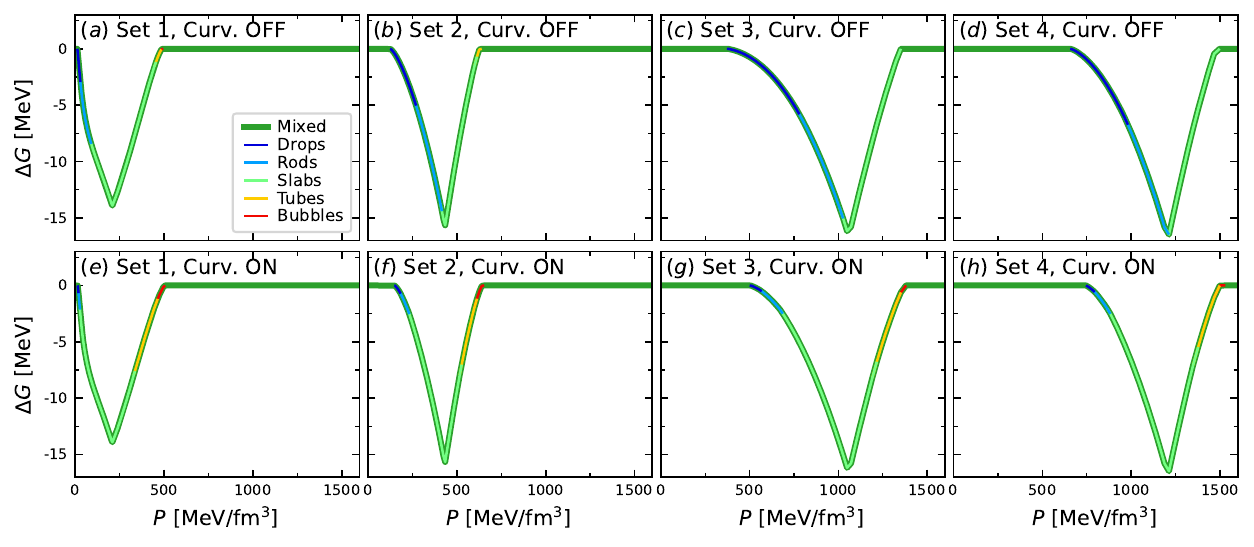}
\caption{$\Delta G$ as a function of pressure in the absence (upper row) and presence (lower row) of curvature effects for the same EoS sets of Fig.~\ref{fig:eos}. The horizontal lines represent the pure hadron and quark phases (hadrons on the left and quarks on the right).  As emphasized in Fig.~\ref{fig:eos}, the absence of curvature favors the formation of drops and rods, while it suppresses the presence of tubes and bubbles. When considering curvature, the Gibbs free energy increases for drops and rods, decreases for tubes and bubbles, and remains unaffected for slabs. As a result, in the lower row there are wider regions featuring tubes and bubbles, while slabs become more prevalent in areas where drops and rods are observed in the upper row.}
\label{fig:delta_gibbs}
\end{figure*}
%-------------------

Fig.~\ref{fig:delta_gibbs_construction} provides a schematic representation of the construction method for the \textit{pasta} phase. This method entails generating both the pure hadronic and quark EoSs, in addition to the EoSs for the five distinct geometric structures. The prevailing phase at a given pressure is the one with the lower Gibbs free energy per baryon, defined as
\begin{equation}
G = \frac{\epsilon + p}{n_B} \, .  
\end{equation}
An alternative approach is to introduce the quantity $\Delta G$, which represents the Gibbs free energy difference between the mixed EoS associated with each geometric structure and the unique Maxwell's construction EoS: 
\begin{equation}
\Delta G_i \equiv  G_{\textrm{Mixed}, i} - G_\textrm{Maxwell} \, ,
\label{eq:delta_G_definition}
\end{equation}
where the index $i$ runs over drops, rods, slabs, tubes and bubbles.
Since the Maxwell's construction EoS remains fixed, the prevailing mixed phase at a given pressure is the one with larger absolute value of $\Delta G$ (see Fig.~\ref{fig:delta_gibbs_construction}). The $\Delta G$ magnitude proves useful for analyzing and distinguishing small energy differences among the various \textit{pasta} phase structures.

The results for the magnitude of $\Delta G$ are presented in Fig.~\ref{fig:delta_gibbs} for the scenarios without (upper row) and with curvature effects (lower row). As in Fig.~\ref{fig:eos},  the dark green envelope curve indicates the whole hybrid mixed EoS and the overlapping colors indicate the prevailing finite-size structure. The energy gain resulting from the formation of geometric structures, in comparison to the Maxwell's construction, falls within the range of a few MeV (specifically, $|\Delta G| < 16 ~\mathrm{MeV}$ as seen in Fig.~\ref{fig:delta_gibbs}). These values can be deemed astrophysically significant, as they are comparable in magnitude to the binding energy of nuclear matter or the binding energy of iron. While drops, rods, tubes, and bubbles are influenced by curvature contributions, slabs remain unaffected by them (cf. Eq.~\eqref{eq:eps_curv} with $d=1$). Consequently, as the absolute minimum value of the $\Delta G$ curves consistently corresponds to slabs, this minimum value remains unaffected when considering the influence of curvature. This figure provides a clearer visualization of the dominant presence of drops, rods and slabs when the curvature effect is not considered. It also exhibits the shift towards a prevalence of slabs and tubes, along with the emergence of bubbles, when curvature is taken into account. Additionally, since the pressure axis in this figure is not logarithmic as in Fig.~\ref{fig:eos}, the shift of the mixed phase range towards higher pressures when altering the EoS set becomes more pronounced and apparent.

%-------------------
\begin{figure*}[tbh]
\centering
\includegraphics[width=.99\linewidth]{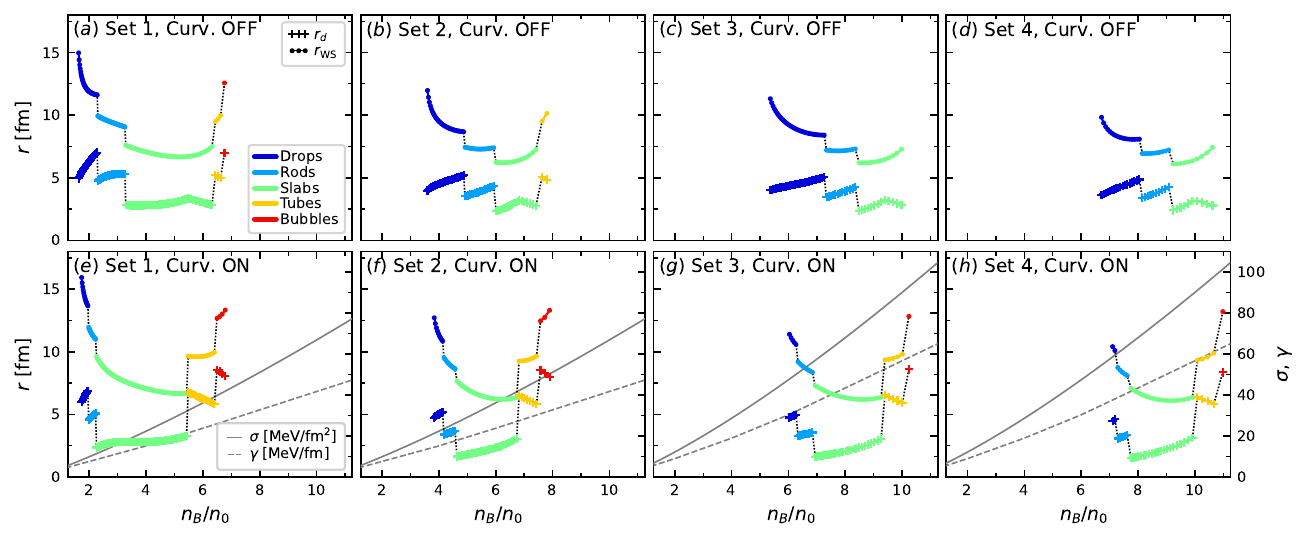}
\caption{Size of the WS cells ($r_\mathrm{WS}$, circle dots) and geometric structures ($r_d$, cross dots) versus $n_B/n_0$, in the absence (upper row) and presence (lower row) of curvature effects. The EoS sets and color coding are consistent with the previous figure. In the lower row, we also show the  dependence of $\sigma$ and $\gamma$ with $n_B/n_0$ (see scale on the right vertical axis of each panel). Specifically, according to the $G_V$ value of each set, panels $(e)$ and $(f)$ display $\sigma$ and $\gamma$ for $G_V=0.2~\mathrm{fm}^2$, while panels $(g)$ and $(h)$ show the corresponding values for $G_V=0.3~\mathrm{fm}^2$.}
\label{fig:radios}
\end{figure*}
%-------------------

%-------------------
\begin{figure*}[tbh]
\centering
\includegraphics[width=.99\linewidth]{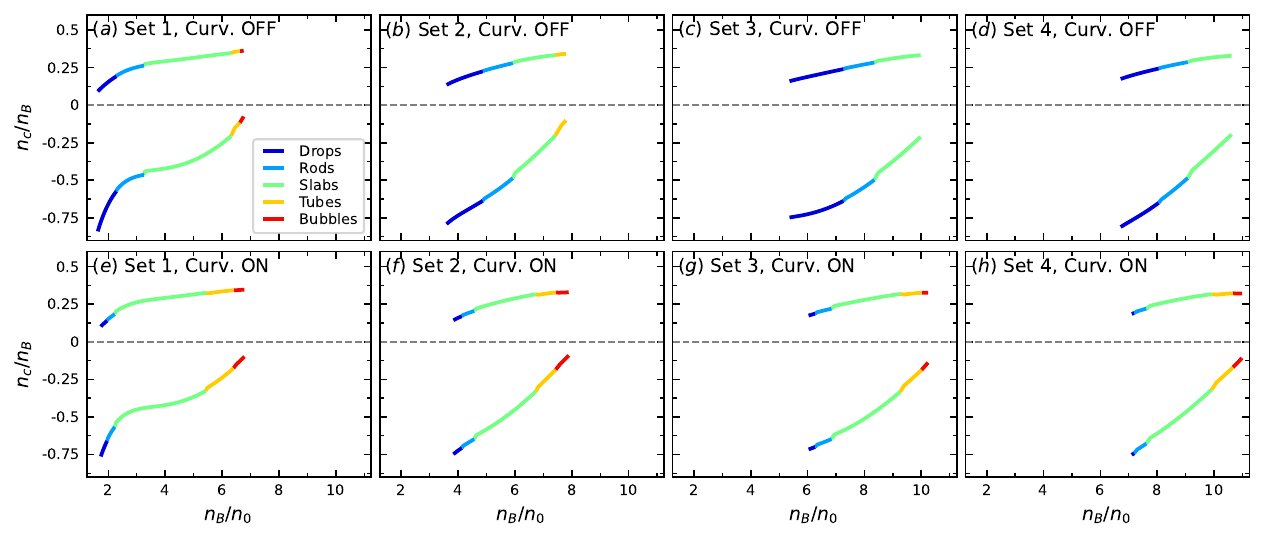}
\caption{Electric charge per baryon $n_c/n_B$ of the hadronic and quark components in the mixed phase, plotted against $n_B/n_0$. The upper row depicts the results in the absence of curvature effects, while the lower row takes into account curvature. The EoS sets and color coding are consistent with previous figures. Positively charged curves represent the hadron phase, and negatively charged curves represent the quark phase.}
\label{fig:nq}
\end{figure*}
%-------------------

%-------------------
\begin{figure*}[tbh]
\centering
\includegraphics[width=0.75\linewidth]{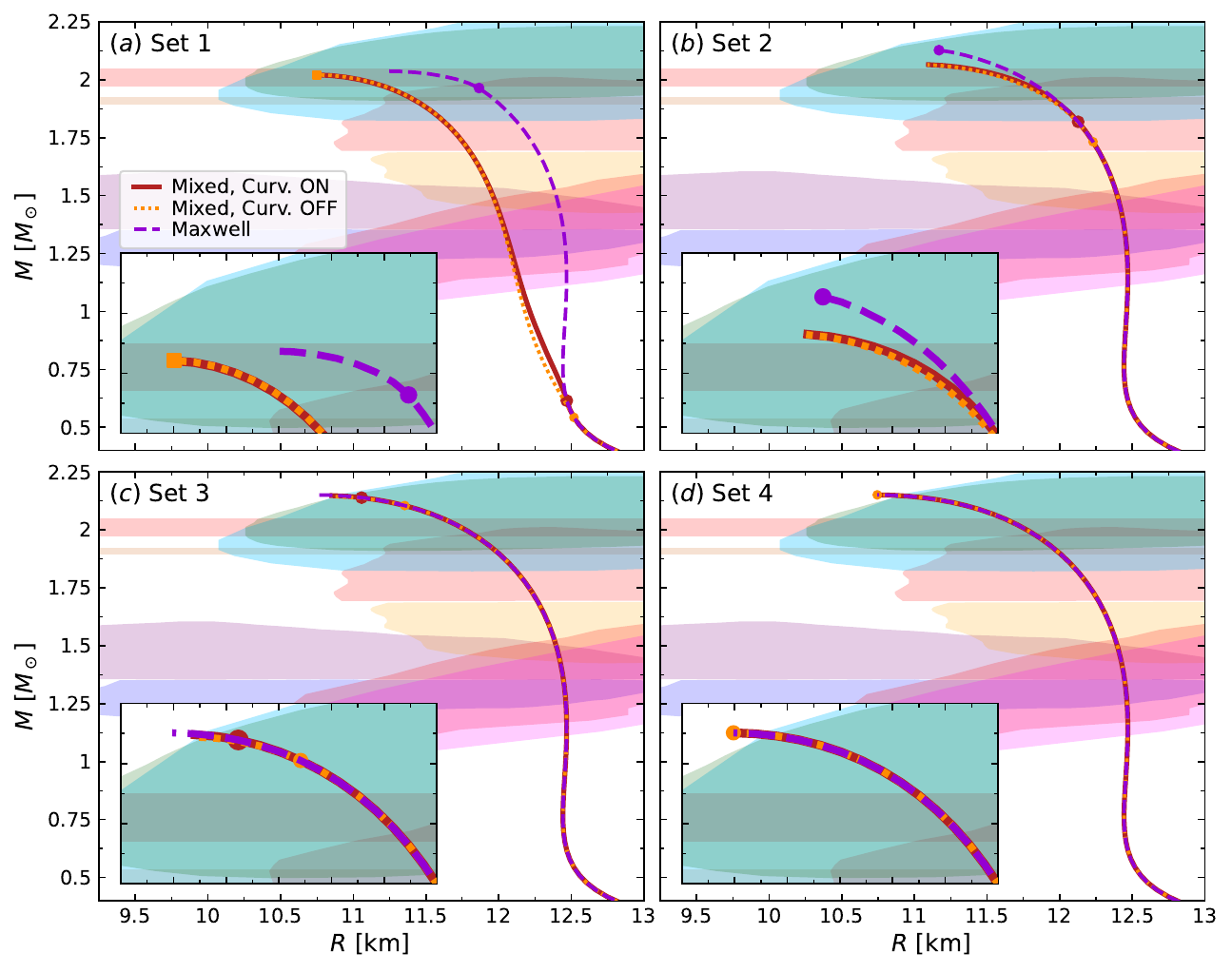}
\caption{Mass-radius relationship for the mixed EoSs with and without curvature effects (solid red and orange dashed curves, respectively), and the Maxwell's construction EoS (dashed violet curve). Each panel corresponds to a different EoS set. The dots over each mixed EoS curve indicate the onset (circle dots) and the end (square dots) of the mixed phase in the stellar core. Circle dots over the Maxwell EoS curves represent the onset of the pure quark core. All EoS sets satisfy current astrophysical constraints (shown by the colored regions). In the zoom box, we provide a detailed view of the maximum mass region. It is noteworthy that although curvature effects modify the variety of geometrical structures within the mixed phase, they do not significantly impact the mass-radius curves. Furthermore, curves of sets~$1$ and $2$ deviate from Maxwell's construction, with the former shifting to smaller radii due to the onset of the mixed phase at very low masses, and the latter showing a slight shift towards smaller radii and a moderate decrease in the maximum mass due to the onset of the mixed phase at a larger mass. In contrast, for sets~$3$ and $4$, the mixed phase emerges above $2~M_{\odot}$, resulting in larger maximum masses with negligible differences between the mixed and Maxwell's construction curves.}
\label{fig:mr}
\end{figure*}
%-------------------

%-------------------
\begin{figure*}[tbh]
\centering
\includegraphics[width=0.99\linewidth]{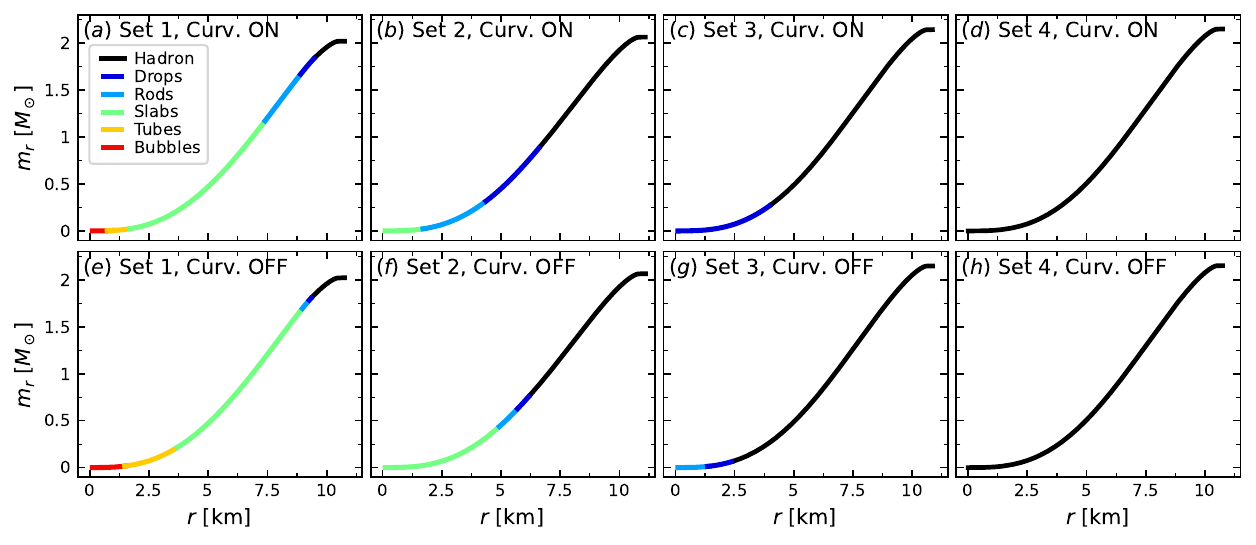}
\caption{
{
Stellar profiles for maximum mass configurations ($m_r$ versus $r$) shown in the absence (upper row) and presence (lower row) of curvature effects. Each panel represents a different EoS set. The black curve indicates the crust and the purely hadronic outer core, while the colored curves represent the dominant finite-size structures in the mixed phase. Curvature effects favor the appearance of slabs, tubes, and bubbles, and reduce the radial extent of the mixed phase within the star. Only the stellar configuration of Set 1 displays the entire variety of the \textit{pasta} phase in its core. No set, for stable objects, reaches a central energy density sufficient to support pure quark matter in its core.
}}
\label{fig:perfil}
\end{figure*}
%-------------------

In Fig.~\ref{fig:radios}, the size of the WS cells and the geometric structures are shown as a function of the baryon number density, for the cases without (upper row) and with (lower row) curvature effects. As previously mentioned in the discussion of Fig.~\ref{fig:eos}, the inclusion of curvature shifts the range of prevalence of the mixed phase towards higher density values. It also alters the density interval in which each geometric structure is energetically favored. Additionally, Fig.~\ref{fig:radios} reveals a slight increase in the sizes of the WS cells, as well as the sizes of drops, rods, tubes, and bubbles for a given baryon density when accounting for curvature effects. 

In the bottom row of Fig.~\ref{fig:radios}, we also show the surface and curvature tensions as functions of the baryon number density $n_B$. As mentioned in Sec. \ref{sec:surf_and_curv}, these curves were obtained from the calculations of Ref. \cite{Lugones:2021tee}. These curves are visually represented in gray and are quantified on the right vertical axis of the corresponding panels. The value of  $G_V$ used in the computation of $\sigma$ and $\gamma$ aligns with that utilized for obtaining the quark EOS. Specifically, we employed $G_V=0.2$~fm$^2$ for sets $1$ and $2$, and $G_V=0.3$~fm$^2$ for sets $3$ and $4$. Due to this factor, the curves featured in panels (e) and (f) are different to the curves displayed in panels (g) and (h). The simultaneous display of $r_d$, $\sigma$ and $\gamma$ enables the identification of the prevailing tension values within each geometric configuration. This presentation also facilitates the estimation of curvature and surface energy densities, as outlined in Eqs. \eqref{eq:eps_surf} and \eqref{eq:eps_curv}.
Notice that the fact that $\sigma$ is an increasing function of density means that the energy cost of forming tubes and bubbles is significantly higher than for forming drops and rods. Therefore, in the top row of Fig. \ref{fig:radios}, tubes and bubbles scarcely appear, while drops and rods form easily. The situation changes qualitatively when curvature effects are included. The sign change observed in $\epsilon_\mathrm{curv}$ in Eq. \eqref{eq:eps_curv} partially counteracts the surface energy cost of  tubes and bubbles, while significantly increasing that of drops and rods. For this reason, drops and rods become less prevalent, and tubes and bubbles emerge more readily, even with higher $\sigma$ and $\gamma$. It is also observed that tubes and bubbles form with significantly larger radii than drops and rods, aiming to reduce the surface and curvature energy.

In Fig.~\ref{fig:nq}, we depict the electric charge per baryon $n_c/n_B$ of the mixed phase for the hadron and quark components as a function of $n_B$. The upper row represents cases without curvature effects, while the lower row shows results that account for curvature. Across all panels, the curve with positive charge corresponds to the hadron component, whereas the negatively charged curve represents the quark component.  Notice that as $\chi$ grows and $ 1-\chi $ decreases, the absolute value of the hadron phase electric charge consistently rises, while that of the quark phase reduces. This behavior ensures mutual compensation to uphold global electric charge neutrality as expressed by Eq.~\eqref{eq:charge}.  

Finally,  we integrated the TOV equations for all the EoSs shown in previous figures.  In Fig.~\ref{fig:mr}, we show the resulting mass-radius relationships providing a comparison between the mixed EoSs with and without curvature effects, as well as the Maxwell's construction hybrid EoS. On the mixed phase curves, two types of markers are used to demarcate transitions: a circular dot symbolizes the emergence of the mixed phase, and a square dot points to the end of the mixed phase and the start of the pure quark phase. In contrast, for the Maxwell EoS curves, a single circular dot is used, marking the phase transition and the onset of the pure quark core. Within the zoom boxes, we provide a detailed view of the maximum mass region. Additionally, color-coded clouds and bars represent the current astrophysical constraints. As previously mentioned, only the results corresponding to EoSs that satisfy these astrophysical constraints are displayed.

Set~$1$ shows the most significant differences between its three curves. The early onset of the mixed phase  (indicated by red and orange circle dots) causes the mixed EoSs curves to deviate from the pure hadronic branch of the Maxwell EoS in the low mass regime. For set~$1$, there are some minimal differences in the EoSs with and without curvature effects for low mass configurations. However, in other mass ranges and for all configurations of the other sets, there is essentially no difference at all between the two scenarios in the mass-radius plane. Notice that only set~$1$ in the case without curvature effects displays a very short segment of stable configurations featuring a pure quark-matter core, as indicated by the presence of a square dot on its curve. This case is an exception. All other stable stellar configurations consist of either only hadronic matter or exhibit a mixed phase core. Additionally, in all cases, the inclusion of curvature energy leads to a delayed appearance of the mixed phase.

For all EoS sets, the incorporation of the \textit{pasta} phase leads to a softening of the EoS. In sets~$1$ and $2$, this softening notably influences the mass-radius relationship because of the early onset of the mixed phase. The resulting hybrid stars support considerably less mass for a given radius than stars of the Maxwell curve. In contrast, for sets~$3$ and $4$, the late appearance of the mixed phase, at very high densities and pressures, produces almost no differences among the three curves of each panel.
This can be seen by the presence of the circle dot at very high masses (or even the absence of the dot, for the red curve of set~$4$). For these two sets, almost all of the configurations in the stable branch for each curve correspond to pure hadronic stars, or have only a tiny mixed phase inner core. The very small differences among the curves that arise in the zoom box for set~$3$, compared to set~$4$, which shows no differences at all, can also be explained by analyzing the position of the dots over the curves. For set~$3$, the mixed phase appears late, but before the maximum mass configurations. For set~$4$, the mixed phase appears in the core of the stellar configurations once they become unstable. Therefore, before the maximum mass, they are all purely hadronic. 

%-------------------
\begin{figure*}[tbh]
\centering
\includegraphics[width=.75\linewidth]{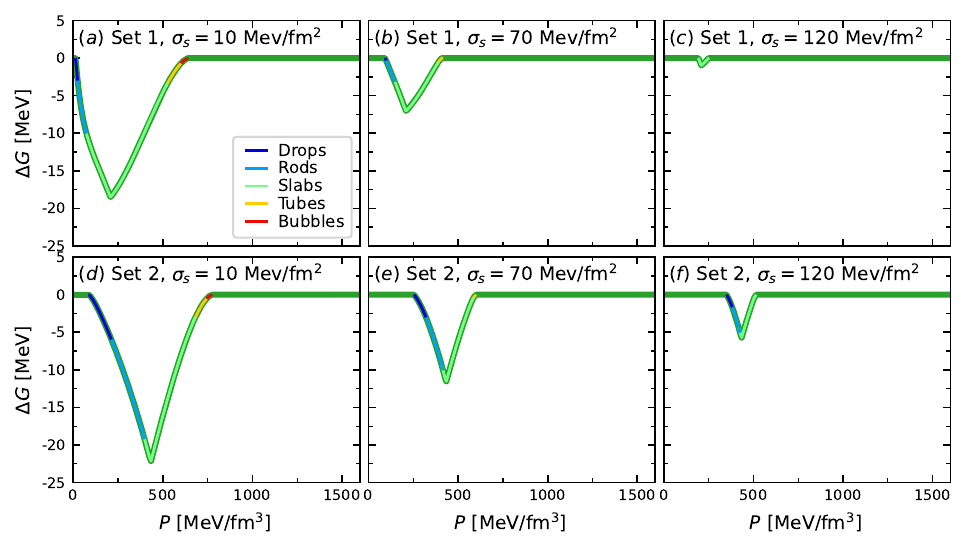}
\caption{Same as in Fig.~\ref{fig:delta_gibbs}, but in the constant surface tension scenario (without curvature effects). Panels $(a)$, $(b)$, and $(c)$ show the results for set~$1$ with $\sigma=10$, $70$, and $120~\mathrm{MeV/fm^2}$, respectively, while panels $(d)$, $(e)$, and $(f)$ display the results for set~$2$ with the same $\sigma$ values. The color coding of the geometric structures remains consistent with Fig.~\ref{fig:delta_gibbs}, and the horizontal lines indicate the pure phases (hadrons on the left and quarks on the right). The mixed EoSs tend to approach the Maxwell's construction EOS as $\sigma$ increases, resulting in $\Delta G$ values that approach zero. Furthermore, as $\sigma$ increases, the variety of geometric structures diminishes: tubes and bubbles shrink, while slab structures become increasingly preferred.}
\label{fig:deltagibbssigmacte}
\end{figure*}
%-------------------

{In Fig.~\ref{fig:perfil}, we present profiles of selected stellar configurations. We depict the $m_r$ versus $r$ relationship, where $r$ represents the radial coordinate of the star, and $m_r$ denotes the gravitational mass of the star integrated up to $r$. Within each panel of the figure, we display the profile of the maximum mass configuration for the respective set, both without (upper row) and with (lower row) curvature effects taken into account. The black curves represent the outer layers, including the crust and the purely hadronic outer core, while the colored curves indicate the prevailing finite-size structures within the mixed-phase core. Firstly, we can observe a correspondence among the EoS results, the $M$-$R$ relationships, and the figure presented here. Set 1, for instance, exhibits the earliest appearance of the mixed phase. Consequently, for the maximum star configuration of Set 1, we observe a larger radial extension. Conversely, Set 4 represents the extreme opposite scenario, where the mixed phase appears only at the very center of the maximum mass configuration, resulting in no mixed phase within the radial range of this star. Note that, since configurations become unstable beyond the maximum mass point, the profiles we present showcase the maximum diversity and extent of \textit{pasta} phase structures attainable within the stable configuration cores for each set. The inclusion of curvature effects, as we have demonstrated, leads to an increased variety of geometrical structures, favoring the prevalence of slabs, tubes, and bubbles. However, it also reduces the radial extent of the mixed phase.  It is noteworthy to mention the significant proportion occupied by the mixed hadron-quark phase, both in terms of radius and mass, in Sets 1 and 2. As previously noted, all sets of stellar configurations become unstable before reaching a central energy density high enough to support pure quark matter in their cores.}

%-----------------------------------------------------------------------------------------------
\subsection{Comparative analysis: medium-dependent versus constant surface tension scenarios}
\label{sec:sigmacte}
%-----------------------------------------------------------------------------------------------

In this subsection, we compare our results, which account for medium-dependent $\sigma$ and $\gamma$, with the widely studied scenario that assumes a constant value for surface tension and neglects curvature effects. To facilitate this comparison, we select three values for the surface tension:
\begin{equation}
\sigma = 10, 70, 120~\mathrm{MeV/fm^2},
\label{eq:cases_sigma_constant}
\end{equation}
and compute results for sets~$1$ and $2$, neglecting curvature effects. 

In Fig.~\ref{fig:deltagibbssigmacte}, we present the relationships between $\Delta G$ and pressure for the chosen values of $\sigma$. As expected,  as $\sigma$ increases, the pressure range where the mixed phase exists decreases for both sets. Additionally, the onset of the mixed phase shifts to higher energy densities. The variation of $\sigma$ also impacts the variety of geometrical structures, with a tendency for diminishing diversity as $\sigma$ increases. At high $\sigma$ values, slabs become the prevalent structure, while in some cases, drops and rods may still coexist. For sufficiently large  $\sigma$, the mixed EoS vanishes, transitioning into a sharp quark-hadron discontinuity represented by the Maxwell's construction. Notably, for set~$2$, the mixed phase converts into a sharp transition for a larger $\sigma$ value compared to set~$1$. 

%-------------------
\begin{figure*}[t]
\centering
\includegraphics[width=0.75\linewidth]{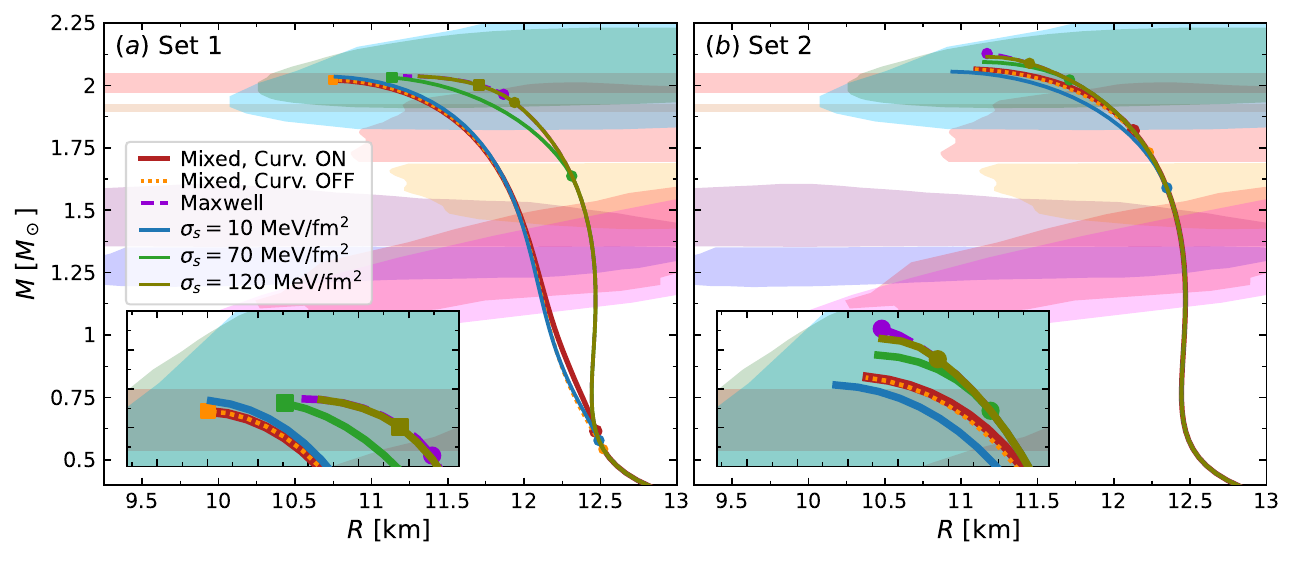}
\caption{Mass-radius relationship for the EoS sets~$1$ and $2$ in the constant surface tension scenario without curvature effects. For comparison, we include the curves from panels (a) and (b) of Fig.~\ref{fig:mr}, which were obtained with medium-dependent $\sigma$. The circle and square dots have the same meanings as in Fig.~\ref{fig:mr}. As $\sigma$ increases, both sets tend to approach the Maxwell's construction curve. For sets $1$ and $2$, the mixed EoS curves with medium-dependent $\sigma$ resemble the curves with $\sigma=10~\mathrm{MeV/fm^2}$ and $\sigma=70~\mathrm{MeV/fm^2}$, respectively.}
\label{fig:mrsigmacte}
\end{figure*}
%-------------------

The results for set~$1$ with medium-dependent $\sigma$ and no curvature effects, as shown in Fig.~\ref{fig:delta_gibbs}, exhibit some similarities to the case with a fixed $\sigma=10~\mathrm{MeV/fm^2}$ in Fig.~\ref{fig:deltagibbssigmacte}. Both cases present a wide mixed phase with all geometric structures present within similar pressure ranges, although the energy gain $\Delta G$  with respect to the Maxwell's construction is smaller in Fig.~\ref{fig:delta_gibbs}. This similarity can be understood by observing the top row of Fig. \ref{fig:radios} and identifying the $\sigma$ values corresponding to each geometric structure in the bottom row. Drops and rods arise at $\sim 2-3 n_0$ with a surface tension around $\sigma \sim 10-20~\mathrm{MeV/fm^2}$, slabs are not affected by surface tension, and the presence of tubes and bubbles is minimal. Therefore, it is unsurprising that the results with density-dependent $\sigma$ are more similar to the case with $\sigma = 10~\mathrm{MeV/fm^2}$ than to the other cases of Eq.~\eqref{eq:cases_sigma_constant}.
Similarly, the results for set~$2$ shown in Fig.~\ref{fig:delta_gibbs} resemble the scenario with a fixed $\sigma=70~\mathrm{MeV/fm^2}$ in Fig.~\ref{fig:deltagibbssigmacte}, with comparable geometric structures emerging in similar pressure ranges, but an even larger difference in $\Delta G$ between both scenarios. In the case of set 2, Fig. \ref{fig:radios} shows that the surface tension across geometric structures varies within the range of $20-60~\mathrm{MeV/fm^2}$. This explains the tendency for these results to bear a closer resemblance to the constant $\sigma$ scenario where $\sigma \sim 70~\mathrm{MeV/fm^2}$.
Finally, when comparing the constant $\sigma$ model with the complete scenario that includes curvature effects (shown in Fig.~\ref{fig:delta_gibbs}), there are fewer similarities between the two cases, primarily because the pressure range covered by the mixed phase is narrower, and the curvature energy favors the emergence of tubes and bubbles in Fig.~\ref{fig:delta_gibbs}.

In Fig.~\ref{fig:mrsigmacte}, we show the mass-radius relationship for various scenarios, including the three constant  $\sigma$ cases and our previous results with medium-dependent $\sigma$ (with and without curvature effects) as well as the Maxwell's construction case. 
In line with our previous figure, the curves exhibiting medium-dependent $\sigma$ are positioned between the cases of $\sigma=10$ and $\sigma=70~\mathrm{MeV/fm^2}$. Set 1 shows a stronger resemblance to the $\sigma=10~\mathrm{MeV/fm^2}$ case, while set~$2$ is more akin to the $\sigma=70~\mathrm{MeV/fm^2}$ case.
The $\sigma=120~\mathrm{MeV/fm^2}$ scenario closely resembles the Maxwell's construction case. In the constant $\sigma$ approximation, the increasing of $\sigma$ stiffens the EoS and shifts the onset of the mixed core to higher densities while the pure quark core emerges at lower densities (as indicated by the phase transition marker points). For set~$1$, the circle and square dots tend to converge as $\sigma$ increases, eventually collapsing into a single dot at the Maxwell's construction limit. On the other hand, for set~$2$, the transition from mixed to pure quark matter occurs in dynamically unstable branches, beyond the maximum mass configurations. Consequently, the curves for set~$2$ gradually modify their maximum mass configurations with varying $\sigma$, while set~$1$ maintains almost the same maximum mass in all scenarios. This difference arises from the fact that the appearance of the mixed phase for set~$1$ occurs for lower mass configurations, making the curves more sensitive to the mixed EoS's softening. Moreover, the softer quark EoS of set~$1$ (with $B=90~\mathrm{MeV/fm^3}$) amplifies this effect compared to the stiffer quark EoS of set~$2$ (with $B=120~\mathrm{MeV/fm^3}$).  Finally, for set~$2$, the mixed phase emerges at higher masses, and the stiffness of the hadronic sector leads to higher maximum mass values. As an extreme case, the Maxwell EoS exhibits the largest maximum mass because the first hybrid object is only possible at the maximum mass configuration.

%-------------------------------------------------------------------
\section{Summary and discussion}
\label{sec:conclus}
%-------------------------------------------------------------------

In this work, we focused on investigating the quark-hadron \textit{pasta} phase in NSs. Unlike most previous analyses that considered the surface tension $\sigma$ as a constant free parameter and neglected the curvature tension $\gamma$, we employed microphysical calculations using the MRE formalism to determine these parameters \cite{Lugones:2021tee}. This approach enabled us to account for the density dependence of both $\sigma$ and $\gamma$ throughout the neutron star, providing a more comprehensive and accurate analysis of the \textit{pasta} phase. To construct the EoSs, we combined an analytic representation based on the APR EoS for hadronic matter with the MIT bag model featuring vector interactions for quark matter. For modeling the mixed phase, we employed the compressible liquid drop model, which accounts for finite-size and Coulomb effects in a consistent manner. We constructed an extensive set of mixed hybrid EoSs by varying model parameters and solved the TOV equations to obtain neutron star mass-radius relationships. Only $6$ out of $18$ initial EoS models met the current astrophysical constraints, with $4$ representative EoSs chosen for detailed analysis.

By construction, the formation of the mixed phase always softens the EoS relative to that derived from the Maxwell's construction. However, the degree of this softening is strongly influenced by the model parameters. For example, when the mixed phase begins at very low pressures, it spans a broad range of densities, leading to pronounced softening in the mixed EoS. Conversely, when the mixed phase occurs at higher densities, its range is more limited, causing its stiffness to approximate that of the Maxwell's construction. Importantly, when considering curvature effects, the EoS's stiffness is largely unaffected. As for the energy gain of the mixed phase over the Maxwell's construction, it largely depends on the type of geometric structure. Nonetheless, the energy difference consistently remains within a few MeV (specifically, $|\Delta G| < 16 ~\mathrm{MeV}$ as seen in Fig.~\ref{fig:delta_gibbs}).

On the other hand, our calculations reveal that the geometric composition of the mixed phase is highly sensitive to the behavior of the parameters $\sigma$ and $\gamma$. Microscopic calculations using  the MRE formalism suggest that both $\sigma$ and  $\gamma$ increase with the rising of the baryon number density. Drops, rods, and slabs form at lower densities, resulting in a lower surface cost compared to tubes and bubbles, which form at much higher densities. Consequently, in the absence of curvature effects, drops, rods, and slabs dominate the mixed phase, while tubes and bubbles are scarcely observed.

The incorporation of curvature introduces significant changes in the geometric composition of the mixed phase. While slabs remain unaffected by curvature, the sign change of $\epsilon_\mathrm{curv}$ in Eq. \eqref{eq:eps_curv} means that surface and curvature effects partially counteract for tubes and bubbles, favoring their existence even at elevated densities and, consequently, higher $\sigma$ and $\gamma$ values. In contrast, for drops and rods, surface and curvature effects combine, making these configurations less predominant. We also find that tubes and bubbles typically form with considerably larger dimensions compared to drops and rods, aiming to minimize both surface and curvature energies.

Our stellar structure calculations suggest that the existence of deconfined quark matter within NSs occurs predominantly within the mixed phase, and, on rare occasions, in a pure quark matter core. The onset of the mixed phase depends  upon the specific EoS adopted, resulting in different NS mass values at which this transition starts. In cases where the onset of the mixed phase occurs at very low masses, the $M-R$ curves that incorporate the mixed phase deviate significantly from the profiles of hybrid stars characterized by sharp quark-hadron interfaces (see curves for sets $1$ and $2$ in Fig. \ref{fig:mr}). Consequently, stars with mixed phases can support considerably less mass for a given radius compared to their counterparts featuring sharp interfaces. These disparities gradually fade as the onset of the mixed phase shifts to higher masses (cf. sets $3$ and $4$ in Fig. \ref{fig:mr}). Remarkably, we observe that the curvature has a minimal impact on the mass-radius relationship, except in scenarios where an extremely low-density onset of the mixed phase is involved, leading to mild effects in those cases.

We compared our results obtained with density-dependent surface tension to those from the more conventional constant surface tension scenario. As expected, our findings revealed a correlation where larger values of $\sigma$ led to stiffer EoSs, culminating in the Maxwell's construction representing the limit of utmost stiffness. Neglecting curvature, the results stemming from the density-dependent surface tension framework could be reasonably approximated by the constant $\sigma$ scenario, with $\sigma$ ranging within $10-70~\mathrm{MeV/fm^2}$, depending on the EoS parametrizations. This similarity extended to the EoS stiffness, the mixed phase's geometrical composition, and the mass-radius relation. However, the inclusion of curvature effects makes it impossible to reproduce the types of geometric structures and their prevalence ranges through a calculation using constant $\sigma$. The primary reason lies in the curvature contribution to the energy density, which is positive for drops and rods but negative for tubes and bubbles. Consequently, there is a counterbalance of finite size energy favoring the formation of tubes and bubbles, while drops and rods tend to be suppressed. This peculiarity defies emulation by a model featuring constant surface tension.

In many prior investigations, the concept of critical surface tension has been explored. Within the framework of a constant $\sigma$, a critical surface tension value, denoted $\sigma_\mathrm{crit}$, emerges \cite{Voskresensky:2002csa}. This threshold value indicates the point at which the energy cost of forming the \textit{pasta} structures becomes prohibitive, as depicted in Fig. \ref{fig:deltagibbssigmacte}. If the (constant) surface tension exceeds $\sigma_\mathrm{crit}$, an abrupt interface between hadrons and quarks is favored. However, our findings indicate that the concept of $\sigma_\mathrm{crit}$ should be employed with caution. As shown in Fig. \ref{fig:radios}, the surface tension in drops and rods is notably lower than in tubes and bubbles. Moreover, as previously discussed, curvature effects favor the formation of tubes and bubbles, even when their associated surface tension is markedly high.

{As discussed in Sec. \ref{sec:modelseos}, employing different quark and hadron EoSs than those chosen for this work would result in variations in the stiffness of the EoS, the quark-hadron transition pressure, and other related factors. However, it is reasonable to anticipate that the fundamental conclusion of our paper would not be qualitatively altered by using a different set of EoSs. As stressed before, the central conclusion of this work is that the inclusion of density-dependent surface tension and curvature tension does not significantly alter the global properties of compact objects but does substantially modify the diversity of geometric structures present in the mixed phase. Our analysis indicates a suppression of drops and rods at low densities and an enhancement of tubes and bubbles at high densities. A change in the EoSs is unlikely to bring major modifications to this trend. For instance, we expect that this general behavior would also be maintained in the context of the Nambu-Jona-Lasinio model with vector interactions, since in this case, both $\sigma$ and $\gamma$ are also increasing functions of density \cite{Villafane:2023sac}. However, it is important to acknowledge that this density-dependent behavior might not necessarily be retained in calculations of $\sigma$ and $\gamma$ using techniques different from the Multiple Reflection Expansion (MRE) formalism \cite{Lugones:2020qll, Lugones:2018qgu, Lugones:2016ytl, Lugones:2013ema, Lugones:2021tee}. }

{Another potentially important aspect is that in our modeling of surface and curvature tensions, we have neglected the contribution from hadronic matter. Generally, it is assumed that quark matter, rather than hadronic matter, provides the dominant contributions to $\sigma$ and $\gamma$, essentially because $\sigma_H$ would be around $1~\mathrm{MeV/fm^2}$ around nuclear saturation density, as suggested by results using semi-empirical mass formulas fitted to observed nuclear masses.  However, detailed calculations estimating the surface and curvature tensions of hadronic matter at densities above nuclear saturation are absent in the literature. Despite these uncertainties, it is conceivable that significant $\sigma$ and $\gamma$ in hadronic matter could  alter the structure of the mixed phase. This stems from the fact that the surface tensions from both phases are additive, and the curvatures are subtractive: $\sigma = \sigma_H + \sigma_Q$ and $\gamma = \gamma_H - \gamma_Q$ \cite{Olesen:1994noq}. If $\sigma_H$ and $\gamma_H$ are sufficiently large, they could substantially increase the effective surface tension and decrease the effective curvature tension. As our results show, both the extent of the mixed phase and the diversity of \textit{pasta} structures are heavily influenced by these quantities.}

In comparison with recent works that also employ the compressible liquid drop model for constructing hybrid stars with a \textit{pasta} phase \cite{Wu:2019zoe, Ju:2021hoy, Ju:2021nev}, it is noteworthy that none of these studies account for curvature effects. While they do not consider curvature, certain qualitative aspects of these investigations align with our findings.   Specifically, there is agreement on the characteristic density range of the mixed phase, the typical sizes of geometric structures, and the expected values of hadron and quark electric charges.   However, a key difference emerges: these studies readily produce all geometric structures (drops, rods, slabs, tubes, and bubbles) within their EoSs, even without considering curvature effects. This contrast can be attributed to their consistent use of relatively low values for $\sigma$, which stand in contrast to the predictions derived from the MRE calculation of surface tension.
Among the previously mentioned works, only one takes into account a density-dependent $\sigma$ \cite{Ju:2021hoy}. Nonetheless, their determination of $\sigma$ using the MRE formalism diverges significantly from our approach. Specifically, they omit the inclusion of an infrared cutoff in their integrals, which is essential to prevent the MRE density of states from becoming negative within a finite volume (as discussed in Refs. \cite{Lugones:2020qll, Lugones:2018qgu, Lugones:2016ytl, Lugones:2013ema, Lugones:2021tee}). Furthermore, their $\sigma$ computation is conducted within the conventional MIT bag model, which implies that their result is the same as for a free Fermi gas of quarks. As a result of this choice, the authors attain markedly low values for $\sigma$ across the entire density range of the mixed phase, specifically $\sigma<50$~MeV/fm$^2$. In contrast, our calculation incorporates both an infrared cutoff and repulsive vector interactions, leading to a quite different behavior of the surface tension.
Additionally, it is worth noting that many EoS models adopted in the aforementioned studies only marginally meet the current astrophysical constraints. In certain cases, the softening effect of the mixed phase in the EoS prevents their models from reaching the $2~M_\odot$ constraint. On the other hand, some studies employ hadronic models that do not agree with the radius ranges implied by the observations from GW170817 \cite{Abbott:2018exr}.

To conclude, our study presents a comprehensive and consistent analysis of the mixed phase in hybrid stars, incorporating a detailed microphysical characterization of two key parameters that influence the properties of the \textit{pasta} phase: surface tension and curvature tension. While our results are consistent with calculations using a constant $\sigma$ in terms of EoS stiffness and the resulting stellar structure, they reveal significant differences in the diversity and range of existence of the geometric configurations. The minor effects on stellar structure due to the density-dependent surface tension and the inclusion of curvature effects should not preclude potential astrophysical implications. For instance, when a hybrid star is dynamically perturbed, the boundary between \textit{pasta} structures and the background might undergo phase conversion reactions. Depending on the variety and extent of these structures, both the effective surface and the conditions facilitating these reactions can vary significantly. The importance of these features cannot be understated, especially considering their potential to affect the damping of $r$-modes \cite{Alford:2010fd} and the dynamic stability of hybrid stars \cite{Pereira:2017rmp, Lugones:2021bkm}.

\section*{Acknowledgments}

M.M. acknowledges CONICET for providing financial support for a research stay at Universidade Federal do ABC, Santo André, SP, Brazil, and for funding through grant PIP-0619. G.L. acknowledges the financial support from the Brazilian agencies CNPq (grant 316844/2021-7) and FAPESP (grant 2022/02341-9).

\bibliography{references}

\end{document}